\documentclass[prb,showpacs,reprint,amsmath,amssymb,superscriptaddress]{revtex4-2}

\usepackage{graphicx}
\usepackage{dcolumn}
\usepackage{bm}
\usepackage{amsthm}
\usepackage{amsmath}
\usepackage{amssymb}
\usepackage{xcolor}
\usepackage{subfig}
\usepackage{caption}
\DeclareCaptionJustification{justified}{\leftskip=0pt \rightskip=0pt \parfillskip=0pt plus 1fil}
\usepackage[export]{adjustbox}
\usepackage{float}
\usepackage{tikz}
\usetikzlibrary{matrix}
\usepackage{braket}
\usepackage{pst-node}

\UseRawInputEncoding

\begin{document}

\preprint{APS/123-QED}
\def\Tr#1{{\rm Tr}\left\{#1\right\}}

\title{Enhancing Spin Coherence of Optically-Addressed Molecular Qubit by Nuclear Spin Hyperpolarization}
\author{Boning Li}
\thanks{These authors contributed equally to this work.}
\affiliation{Department of Physics, Massachusetts Institute of Technology, MA 02139, USA}

\author{Patrick Hautle}
\thanks{These authors contributed equally to this work.}
\affiliation{PSI Center for Neutron and Muon Sciences, CH-5232 Villigen-PSI, Switzerland}

\author{Duhan Zhang}
\thanks{These authors contributed equally to this work.}
\affiliation{Department of Chemistry, University of Pennsylvania, Philadelphia, PA 19104, USA}

\author{Liangping Zhu}
\affiliation{Department of Chemistry, University of Pennsylvania, Philadelphia, PA 19104, USA}
\author{Ashley Beers}
\affiliation{Department of Chemistry, University of Pennsylvania, Philadelphia, PA 19104, USA}
\author{Zeyu Wang}
\affiliation{Department of Chemistry, University of Pennsylvania, Philadelphia, PA 19104, USA}

\author{Paola Cappellaro}\email[Corresponding author: ]{pcappell@mit.edu}
\affiliation{Department of Physics, Massachusetts Institute of Technology, MA 02139, USA}
\affiliation{Department of Nuclear Science and Engineering, Massachusetts Institute of Technology, Cambridge, MA 02139, USA}

\author{Tom Wenckebach}\email[Corresponding author: ]{tom@wenckebach.net}
\affiliation{PSI Center for Neutron and Muon Sciences, CH-5232 Villigen-PSI, Switzerland}

\author{Yifan Quan}\email[Corresponding author: ]{yquan@sas.upenn.edu}
\affiliation{Department of Chemistry, University of Pennsylvania, Philadelphia, PA 19104, USA}
\date{\today}

\begin{abstract}
Optically addressable molecular triplet spins provide a chemically tunable platform for quantum application, but their coherence is often limited by interactions with surrounding spin baths. Here we demonstrate controlled suppression of nuclear-bath-induced decoherence in photoexcited triplet spins of pentacene co-crystallized in high-purity naphthalene single crystals. By hyperpolarizing the proton spin bath through triplet dynamic nuclear polarization (triplet-DNP), magnetic noise generated by the nuclear spins is suppressed, leading to an extension of the electron spin transverse coherence time. Experimentally, we observe a 25\% enhancement of the spin-echo decay time with $60\%$ polarization of the proton spin bath. The measured scaling of the spin-echo decay time ($T_2$) with nuclear polarization quantitatively follows the predicted dependence derived from the polarization-controlled nuclear second moment. 
Both the enhancement and the absolute value of the coherence time are quantitatively reproduced by cluster correlation expansion (CCE) simulations.

These results establish nuclear spin hyperpolarization as a general and actively tunable approach to engineering coherence in molecular qubits. This work provides a broadly applicable design framework for high-coherence molecular and solid-state spin systems.

\end{abstract}


\maketitle

\section{Introduction}

Spin-based qubit platforms have been widely explored for quantum computing~\cite{ladd2010quantum,abobeih2022fault,loss1998quantum}, study of many-body quantum dynamics~\cite{choi2017observation,kucsko2018critical,gao2025dressed}, and quantum sensing~\cite{degen2017quantum,li2025robust,wang2022sensing} over the past decade. In addition to well-established solid-state defect qubits, organic molecular electron spins with low spin-orbit coupling have recently emerged as a promising platform~\cite{Kohler1993, Wrachtrup1993}. These molecular qubits can be optically initialized, exhibit long coherence times, and allow efficient manipulation due to the strong interaction between electron spins and external electromagnetic fields~\cite{Singh2025, Mena2024,singh2025high}. 

Moreover, the electron spins in the molecule are typically surrounded by addressable nuclear spins that can form versatile multi-spin quantum registers for local quantum memory, computation and sensitive detection of local electromagnetic environments. Chemical tunability further provides a powerful route to functional molecular design and scalable synthesis~\cite{Mann2025,Sakamoto2023,Avalos2020}, allowing precise control of the spin environment~\cite{Singh2025, Mena2024,Mann2025,yamauchi2024modulation} and facilitating device integration~\cite{Zadrozny2017, Bayliss2020, Tateishi2026}.

For broad quantum applications of molecular spin qubits, preserving coherence remains a central challenge. Recent studies have shown that intermolecular electron-electron interactions can be effectively suppressed at low spin concentrations. However, decoherence arising from the surrounding nuclear spin bath remains significant, particularly in organic systems where abundant hydrogen nuclei form a dense proton bath. 
Various strategies have been developed to mitigate nuclear-spin-induced decoherence, including dynamical decoupling techniques that suppress low-frequency noise~\cite{li2025robust,Aiello2013,hirose2012continuous,Solomon1959,khodjasteh2005fault,zhang2007dynamical}, nuclear spin preparation methods that reduce Overhauser field fluctuations~\cite{reilly2008suppressing,imamoglu2003optical,ramon2007dynamical,witzel2008wavefunction}, and chemical approaches such as ligand engineering or isotopic substitution (e.g., deuteration) that modify the spin environment~\cite{Graham2017,Graham2017Forging,wedge2012chemical,Ardavan2007,Morton2006,ishiwata2025,balasubramanian2009ultralong}.

Here, we demonstrate a sizable increase in the coherence time of a pentacene electronic spins by hyperpolarizing the proton spin bath. Pentacene is a particularly attractive platform for its long-lived ($\sim100~\mu$s), optically addressable electronic triplet state, which is highly spin polarized ($>90\%$) by optical excitation. These properties make pentacene co-crystallized in a host matrix a promising qubit platform  \cite{li2025robust, Singh2025, Mena2024, Kohler1993,Wrachtrup1993,ishiwata2025,Moro2022,Lin1997, Yunus2022,zhang2024ultra,wu2022enhanced}, including our recent demonstration of vector AC field sensing \cite{li2025robust}.
Important for our demonstration, this high electron polarization can moreover subsequently be transferred to nearby nuclear spins via dynamic nuclear polarization (DNP)~\cite{Hautle2024,Quan2022,Quan2019,Steiner2023,Tateishi2026,Kouno2019,Nishimura2019,Eichhorn2014,Eichhorn2022,Tateishi2014}. In contrast to molecules with a paramagnetic ground state, the triplet spin of pentacene decays to a spinless singlet ground state, allowing the pre-hyperpolarized proton bath to retain its polarization for a much longer time ($\sim 800$ h~\cite{Quan2019}). As we will demonstrate, the polarized nuclear bath will provide a reduced magnetic noise environment leading to an enhanced electron spin coherence, even during many repeated electron spin triplet excitations.

\begin{figure*}[htbp]
\includegraphics[width=1\textwidth]{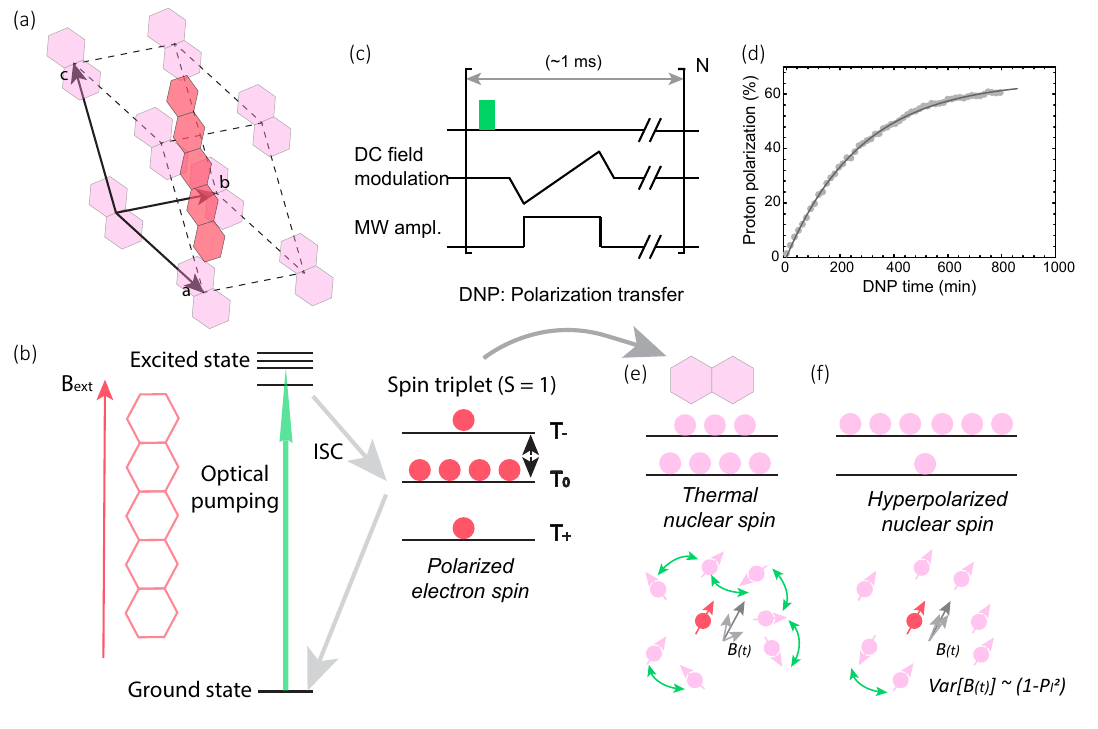}
\caption{\textbf{Pentacene triplet spin system and proton spin bath hyperpolarization.} 
(a) Lattice structure of naphthalene (light red) with a pentacene molecule substitution (red). The nuclear spin bath is dominated by protons in the protonated naphthalene host. The pentacene molecular electron spin can be optically excited and interacts with the surrounding nuclear spins through magnetic dipolar coupling.
(b) Molecular structure of pentacene and spin energy levels. The experiment is performed with an external magnetic field applied along the long molecular axis of pentacene ($\sim$0.355~T). Optical excitation by a laser pulse and spin-selective intersystem crossing (ISC) populates the photo-induced triplet manifold ($T_+,T_0,T_-$), generating a highly spin-polarized electron ensemble. 
(c) Field-swept integrated solid effect (ISE) sequence used for triplet dynamic nuclear polarization (triplet-DNP). 
During the transient triplet lifetime ($\sim 100~\mu$s), the initialized electron spin polarization can be transferred to nearby proton spins via microwave irradiation ($\sim7\,\mu$s) while sweeping the magnetic field over a range of $\sim 2$~mT. This sequence is repeated at a kHz repetition rate to accumulate polarization in the nuclear spin bath.
(d) Experimental build-up of nuclear spin polarization as a function of the DNP accumulation time. The polarization is obtained from the measured proton nuclear magnetic resonance (NMR) signal.
(e,f) Schematic illustration of the proton nuclear spin bath before and after DNP. In thermal equilibrium the nuclear spins are nearly unpolarized, producing fluctuating magnetic fields at the electron spin. Partial polarization of the nuclear spins reduces these magnetic field variance and fluctuations, thereby suppressing electron spin decoherence.
\label{fig:scheme_triplet}
}
\end{figure*}

\section{Results and Discussions}
\begin{figure*}[htbp]
\includegraphics[width=1\textwidth]{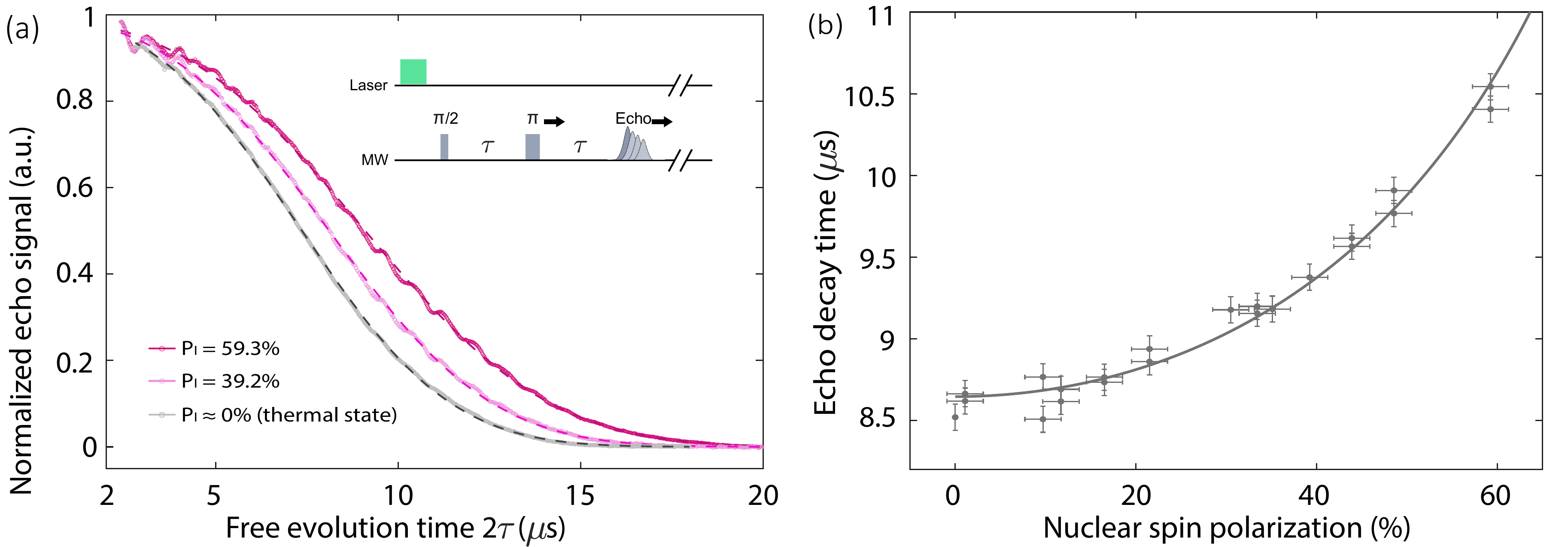}
\caption{
\textbf{Triplet decoherence measurements.} 
(a) Normalized echo signal intensity as a function of the total evolution time $2\tau$ for different nuclear polarizations $P_I$. The decay is fitted to a stretched exponential function,
$E(2\tau)=\exp\!\left[-(2\tau/T_2)^\mu\right]$.
Increasing nuclear polarization leads to a slower decay of the echo signal, demonstrating that nuclear hyperpolarization protects the coherence.
(b) Extracted coherence time $T_2$ as a function of nuclear polarization $P_I$. A 25\% enhanced electron coherence was measured with $P_I=60\%$. A linear function is fitted to $\log T_2$ versus $\log(1-P_I^2)$ (shown as the solid line) and the fitted slope is $-0.46$.
\label{fig:T2_sum}
}
\end{figure*}
Figure~\ref{fig:scheme_triplet}  (a,b) shows the pentacene-naphthalene crystal structure, and a detailed energy level diagram for optically excited triplet states in pentacene with an external magnetic field~\cite{Quan2019,Quan2023}. Upon excitation with a short laser pulse, the molecule is excited from the singlet ground state to the excited singlet state $S_1$, followed by intersystem crossing (ISC) into  a metastable triplet manifold $T_{+},\,T_0\,,T_-$ (lifetime $\sim$100~$\mu$s). ISC is spin dependent and preferentially populates the $T_0$ triplet sublevel,  with $\rho_{0} \sim 0.91 \gg \rho_+,\rho_-$ resulting in a highly polarized electron spin ensemble~\cite{van1980epr}. 

To access the transient triplet state of pentacene, time-resolved electron paramagnetic resonance (TR-EPR) measurements were performed using a home-built X-band pulsed EPR spectrometer, with microwave pulses and signal detection synchronized to photo-excitation by a pulsed laser system operating at 515~nm \cite{Eichhorn2013}. An external magnetic field ($\sim$0.355~T) is applied along the long axis of the pentacene molecule, and a microwave field at 9.44~GHz drives the $T_0 \leftrightarrow T_-$ high field transition (for detailed experimental information, see the Supporting Information).

Deuterated pentacene co-crystallized in a naphthalene single crystal is studied. At a low doping concentration ($\sim 0.7~\mathrm{mM}$), electron spin-spin interactions are suppressed, and the coherence is primarily limited by the coupling between the electron spin and the proton nuclear spin bath in the naphthalene host~\cite{li2025exploring,li2025robust}.

The interaction between the electron spin and the surrounding nuclear spin bath can be described by the Hamiltonian
\begin{equation}
{\cal H} = S_z \sum_{j}\sum_{\sigma = z,\perp} A_{z\sigma}^{(j)} I^{(j)}_{\sigma}(t),
\label{eq:hamiltonian_int_initial}
\end{equation}
where $S_z$ is the electron spin operator along the quantization axis defined by the external magnetic field, and $I_{\sigma}^{(j)}(t)$ denotes the $\sigma$-component of the $j$-th nuclear spin operator.

When the electron spin is prepared in a superposition state couplings to the (time-dependent) nuclear spins  lead to random phase accumulation of the electron spin, thus limiting the transverse coherence.
In this work, we extend the spin-echo coherence time by hyperpolarizing the nuclear spin bath.  Intuitively, a polarized nuclear spin bath suppresses nuclear spin flip–flop processes and reduces the randomness of the nuclear spin configuration. As a result, the nuclear spin bath becomes more static and narrow, as illustrated in Fig.~\ref{fig:scheme_triplet}(e,f), allowing the accumulated phase from the interaction described by Eq.~\eqref{eq:hamiltonian_int_initial} to be more effectively refocused by the spin-echo pulse.

Experimentally, nuclear spin bath polarization is achieved using triplet-DNP~\cite{Hautle2024,Quan2022,Steiner2023,Tateishi2026,Kouno2019,Nishimura2019,Eichhorn2014,Eichhorn2022,Tateishi2014} via the field-swept integrated solid effect (ISE)~\cite{Henstra1990,Henstra2014,Eichhorn2014,Quan2022,Hautle2024, Quan2022,Quan2023b,Can2018} (Fig.~\ref{fig:scheme_triplet} (c)). In each DNP cycle, the electron triplet state is photo-excited and its polarization transferred to the nuclear spins. This process is repeated and the accumulating nuclear spin polarization can be measured by nuclear magnetic resonance (NMR), see Fig.~\ref{fig:scheme_triplet}(d), (see Supplementary Information for experimental details).

Following nuclear spin hyperpolarization, the  coherence time of the electron spin, $T_2$, was measured at 80 K sweeping the inter-pulse delay , $\tau$, of a Hahn-echo sequence. Further details of the experimental implementation are provided in the Supplementary Information.
Figure~2 shows the electron spin decoherence measured under different nuclear spin polarization conditions. 
Consistent with the intuitive picture introduced above, we observe a systematic increase in the electron spin coherence time with increasing nuclear spin polarization.

The Hahn-echo decay in Fig.~\ref{fig:T2_sum}(a) is fitted to a stretched-exponential form,
\(E(2\tau)=\exp\!\left[-(2\tau/T_2)^\mu\right].\)
For all datasets, the extracted stretching exponent $\mu$ lies between 2.5 and 2.8, indicating a slowly fluctuating environment characteristic of a nuclear spin bath. This behavior can be phenomenologically captured by modeling the electron spin as subject to a classical stochastic magnetic field $B(t)$ described by an Ornstein-Uhlenbeck (OU) process arising from weak nuclear spin-spin interactions and nuclear spin-lattice relaxation~\cite{cywinski2008enhance,hall2014analytic}. The interaction Hamiltonian (Eq.~\eqref{eq:hamiltonian_int_initial}) can be rewritten as
\begin{equation}
{\cal H}(t)=\gamma S_z B(t),
\label{eq:hamiltonian_int_classical}
\end{equation}
where $\gamma$ is the gyromagnetic ratio of the electron spin and $B(t)=\frac1\gamma \langle\sum_j \sum_{\sigma} A_{z\sigma}^{(j)} I_\sigma^{(j)}(t)\rangle$ is treated as a classical field.

For a partially polarized nuclear spin bath described by $\rho^{(j)}=\frac{1}{2}(I^{(j)}+2P_I I_z^{(j)})$,
the Overhauser field can be decomposed into a static component $\langle B\rangle= \sum_j A_{zz}^{(j)} P_I$ and a fluctuating component $\delta B(t)=B(t)-\langle B\rangle$ which arises from nuclear spin flip-flop. The static component is refocused by the Hahn echo and does not contribute to decoherence, while the fluctuating component governs the decay.

Given a stochastic field $\delta B(t)$, the spin-echo coherence is given by
\begin{equation}
\langle S_x(2\tau)\rangle=\exp\left[-\frac{\gamma^2}{4\pi}
\int_0^\infty S(\omega)\,|F(2\omega\tau)|^2 d\omega\right],
\label{eq:decoherence}
\end{equation}
where $S(\omega)=\int dt\, \langle\delta B(t)\delta B(0)\rangle e^{-i\omega t}$ is the noise power spectral density, and $F(2\omega \tau)$ is the Hahn-echo filter function
~\cite{cywinski2008enhance,de2010universal}.

In the OU model, $S(\omega)$ is a Lorentzian  centered at zero frequency, reflecting longitudinal ($I_z$) flips of the nuclear spin bath, with additional components at the nuclear Larmor frequency $\pm\omega_{0I}$ arising from transverse ($I_\perp$) dynamics. Under the experimental conditions ($\omega_{0I} \approx 15$~MHz and $2\tau \in [2,15]~\mu$s), the transverse components at $\omega \sim \omega_{0I}$ lie outside the bandwidth of the Hahn-echo filter function and therefore contribute negligibly to the decoherence in Eq.~\eqref{eq:decoherence}.
The spectrum can therefore be approximated as
\begin{equation}
S(\omega)\approx
\frac{2B_{\mathrm{rms},z}^2\tau_c}{1+\omega^2\tau_c^2},
\label{eq:approx}
\end{equation}
where $B_{\mathrm{rms}}^2\equiv\langle\delta B(0)\delta B(0)\rangle$, and $\tau_c$ is the correlation time of the spin bath.
Spin polarization affects both these parameters.

The variance of the nuclear spin effective longitudinal field is
\begin{equation}
\begin{aligned}
    B_{\mathrm{rms},z}^2 \approx& \sum_j (A_{zz}^{(j)})^2
\left[\langle (I_z^{(j)})^2 \rangle - \langle I_z^{(j)} \rangle^2\right]\\
\propto& \sum_j (A_{zz}^{(j)})^2 (1 - P_I^2),
\end{aligned}
\end{equation}
which indicates how  the amplitude of the fluctuating magnetic field is suppressed by nuclear polarization, in turns leading to a reduced electron spin decoherence rate.

Nuclear spin polarization also affects the spin bath correlation time. At cryogenic temperature where nuclear spin-lattice relaxation is strongly suppressed ($\sim 50$~h~\cite{Steiner2023}), the bath dynamics are dominated by nuclear spin flip-flop processes driven by dipolar interactions. 
The rate of flip-flops is set by the dipole-dipole coupling strength $J$ and the energy difference between spin pairs, $\Delta\omega$, yielding $1/\tau_c\sim J_{\mathrm{rms}}^2/\Delta\omega$. 
As the polarization increases, the probability of two spins flip-flopping decreases with the probability of finding two spins of opposite polarization, $\sim (1-P_I)(1+P_I)$. 
Conversely, the same line-narrowing mechanism that reduces $B_{\mathrm{rms},z}^2$, also leads to an increases the flip-flop rate as more spin-pairs are close to resonance.  Previous studies show that the NMR linewidth narrows with increasing polarization, which can be captured by \cite{Abragam1973}
\begin{equation}
\Delta\omega \propto (1-P_I^2)^{-1/2},
\end{equation}
We can then evaluate the flip-flop rate as $1/\tau_c\sim \frac{J_{\mathrm{rms}}^2(1-P_I^2)}{\Delta\omega\sqrt{1-P_I^2}}$.  (See more details in the supplementary information).

Based on this analysis, we evaluate the electron spin coherence time $T_2$ as a function of the nuclear spin bath polarization. In the long correlation-time limit, $\tau_c \gg 2\tau$, Eq.~\eqref{eq:decoherence} reduces to
\begin{equation}
\langle S_x(2\tau)\rangle \approx
\exp\left[-\left(\frac{\gamma^2 B_{\mathrm{rms}}^2}{12\tau_c}\right)(2\tau)^3\right],
\label{eq:t2_large_tauc}
\end{equation}
which yields \(T_2 \propto \left(\frac{\tau_c}{\gamma^2 B_{\mathrm{rms}}^2}\right)^{1/3}\). Therefore, we obtain
\begin{equation}
T_2 \propto (1-P_I^2)^{-1/2}.
\end{equation}
This prediction is consistent with the experimental results, where a linear fit of $\log T_2$ versus $\log(1-P_I^2)$ yields a slope of approximately $-0.46$, shown in Fig.~\ref{fig:T2_sum}~(b).

Under our experimental conditions, the inter-nuclear coupling strength is $J_{\mathrm{rms}}\sim 30$~kHz, corresponding to a correlation time $\tau_c\sim 30~\mu$s. As a result, the condition $\tau_c \gg 2\tau$ is not always well satisfied, which can lead to a stretch exponent smaller than 3.  In addition, the proton spin bath is not spatially homogeneous: protons within the same naphthalene molecule exhibit faster flip-flop dynamics than those in different molecules, making a single effective correlation time only approximate. Moreover, interactions among nuclear spins lead to spatially inhomogeneous energy shifts that modifies the flip-flop rate. Finally, for protons located close to the electron spin (``frozen core''~\cite{khutsishvili1966spin}), the dynamics can be influenced by the coupling to central spin and are not fully captured by a classical noise description, forming a ``quantum'' spin bath. 
\begin{figure}[htbp]
\centering
\includegraphics[width=0.5\textwidth]{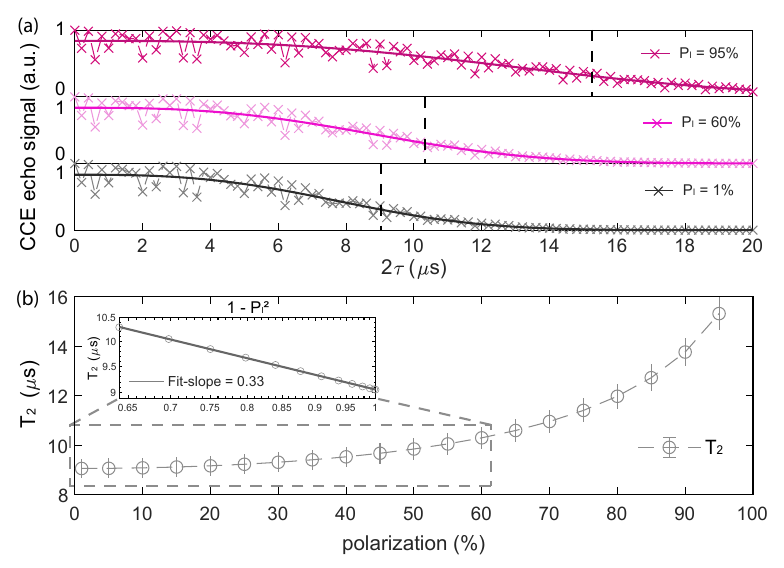}
\caption{\textbf{CCE-calculated results.} 
(a) Hahn-echo decay simulated using CCE at different nuclear spin polarizations, assuming the electron spin is localized at the center of the pentacene molecule. Oscillations arise from strongly coupled proton spins in the naphthalene lattice (Fig.~\ref{fig:scheme_triplet}(a)). 
(b) Extracted $T_2$ from stretched-exponential fits. Inset: $T_2$ versus $(1-P_I^2)$ on a log-log scale; for $p_I<60\%$, a linear fit yields a slope of $\sim 0.4$, consistent with experiment.}
\label{fig:CCE}
\end{figure}

To account for both nuclear-nuclear interactions and the quantum-classical spin bath, we performed cluster correlation expansion (CCE) simulations of the proton bath in the naphthalene lattice which directly simulating the many-body dynamics~\cite{Yang2008,Onizhuk2021,Jahn2022,Jahn2024,li2025exploring}. Details of the simulations are provided in the Supplementary Information.

Figure~\ref{fig:CCE} shows the CCE-simulated Hahn-echo decay of the electron spin at different nuclear spin polarizations, together with the extracted $T_2$ values obtained from stretched-exponential fits. Quantitatively, the decay rate extracted from the CCE simulations is smaller than the experimentally measured decay rate by approximately 5~\%. This difference is consistent with the estimated contribution from electron-electron spin interactions in pentacene and the triplet lifetime (see Supporting Information for details). The good quantitative agreement between experiment and simulation supports the conclusion that the system is relatively clean, and that nuclear spins constitute the dominant source of decoherence.

CCE simulations predict that the electron spin coherence time $T_2$ could reach $\sim 16~\mu\mathrm{s}$ at nuclear spin polarizations up to $95\%$, which is close to the theoretical limit set by photoexcited triplet polarization. This corresponds to approximately a twofold enhancement compared to the unpolarized thermal nuclear spin bath. Experimentally, a record proton spin polarization of $80\%$ has been achieved using a higher-power 556~nm laser~\cite{Quan2019}. In the simulations presented in Figure~\ref{fig:CCE}, the electron spin is assumed to be localized at the center of the pentacene molecule. In the Supplementary Information, we also present ensemble-averaged simulation results that account for the delocalization of the electron spin over the pentacene molecule~\cite{sakamoto2023polarizing,richert2017delocalisation}. This spatial delocalization modifies the hyperfine couplings to the surrounding nuclear spin bath due to their distance dependence~\cite{Sloop1990}. However, the overall decoherence remains largely unaffected, as it is dominated by the large number of weakly coupled, distant nuclear spins.

In summary, we have experimentally demonstrated that nuclear spin hyperpolarization provides a practical and effective route to enhancing the coherence of optically addressable molecular qubits. A 25\% enhancement of the electron transverse coherence time was experimentally achieved with 60\% nuclear spin bath polarization. Combined with analytical modeling and CCE simulations, our results establish a comprehensive understanding of the underlying decoherence mechanisms in the optically addressable pentacene triplet system, which are governed predominantly by coupling to the surrounding nuclear spin bath. Quantitative agreement is achieved between the experimental data and both the analytical theory and the CCE simulations on the absolute timescale.

Building on these results, and leveraging the exceptionally long nuclear spin polarization lifetimes in ultrapure naphthalene crystals~\cite{Quan2019}, we propose a sensing protocol in which the nuclear spin bath is first hyperpolarized to enhance the electron spin coherence time, after which the electron spin serves as the sensing element. Importantly, nuclear spin initialization is required only once, as the polarization persists on timescales far exceeding typical sensing sequences. In the absence of optical excitation, the nuclear $T_1$ exceeds $50$~h at 80~K~\cite{Steiner2023} and 800~h at 25~K~\cite{Quan2021}. Also leveraging the long relaxation time, such highly polarized samples can be transported without compromising the nuclear polarization, enabling spin preparation at one location and sensing at another~\cite{Steiner2023}. During sensing, optical excitation and microwave control of the electron spin induce additional nuclear depolarization. Experimentally, we observe only $\sim 2\%$ nuclear spin polarization loss after 30~min of continuous operation using a Hahn-echo sequence at a 100~Hz repetition rate (see Supplementary Information). This corresponds to $\sim 9\times10^6$ Hahn echo cycles, enabling stable operation over hours to days while maintaining enhanced spin coherence. A similar approach has been explored in quantum dot systems, where $\sim$1\% nuclear spin polarization leads to suppression of Overhauser field fluctuations lasting on the order of one minute~\cite{reilly2008suppressing,foletti2009universal}.

Importantly, the results of this work extend beyond pentacene. Nuclear spin hyperpolarization provides a broadly applicable strategy for enhancing coherence in molecular qubits and other solid-state spin systems. This approach can be integrated with existing coherence-protection techniques, such as dynamical decoupling control~\cite{li2025robust,Aiello2013, hirose2012continuous,Solomon1959}, as well as chemical strategies such as isotopic substitution~\cite{Graham2017,Graham2017Forging,wedge2012chemical,ishiwata2025,balasubramanian2009ultralong}. More generally, these results provide a framework for the rational design of scalable molecular spin-based quantum devices.

\section*{Acknowledgement}
\textbf{Funding:} B. L. thanks MathWorks for their support in the form of a Graduate Student Fellowship. The opinions and views expressed in this publication are from the authors and not necessarily from MathWorks.
\textbf{Author contributions:} Y.Q., B.L., T.W., and P.H. conceived the idea, planned the study, and designed the experiments. P.H. carried out experiments. Y.Q., D.Z., B.L., L.Z., A.B., and Z.W. analyzed the data. B.L., Y.Q., and P.C.  carried out the theoretical analysis and CCE simulations. Y.Q. and P.H. were responsible for pentacene-naphthalene synthesis. Y.Q., T.W. and P.C. supervised the project. All authors discussed the results, and wrote the manuscript. 
\textbf{Competing interests:} The authors declare that they have no
competing interests. \textbf{Data and materials availability:} All data needed to evaluate the conclusions
in the paper are present in the paper and/or the Supplementary Materials. Additional data
related to this paper may be requested from the authors
\bibliographystyle{apsrev4-2}
\bibliography{PentaceneT2_NuclearPolarization_ref}

\end{document}


\title{Enhancing Spin Coherence Time of Optically-Addressed Molecular Qubit by Nuclear Spin Hyperpolarization}
\author{Boning Li}
\thanks{These authors contributed equally to this work.}
\affiliation{Department of Physics, Massachusetts Institute of Technology, MA 02139, USA}

\author{Patrick Hautle}
\thanks{These authors contributed equally to this work.}
\affiliation{PSI Center for Neutron and Muon Sciences, CH-5232 Villigen-PSI, Switzerland}

\author{Duhan Zhang}
\thanks{These authors contributed equally to this work.}
\affiliation{Department of Chemistry, University of Pennsylvania, Philadelphia, PA 19104, USA}

\author{Liangping Zhu}
\affiliation{Department of Chemistry, University of Pennsylvania, Philadelphia, PA 19104, USA}
\author{Ashley Beers}
\affiliation{Department of Chemistry, University of Pennsylvania, Philadelphia, PA 19104, USA}
\author{Zeyu Wang}
\affiliation{Department of Chemistry, University of Pennsylvania, Philadelphia, PA 19104, USA}

\author{Paola Cappellaro}\email[Corresponding author: ]{pcappell@mit.edu}
\affiliation{Department of Physics, Massachusetts Institute of Technology, MA 02139, USA}
\affiliation{Department of Nuclear Science and Engineering, Massachusetts Institute of Technology, Cambridge, MA 02139, USA}

\author{Tom Wenckebach}\email[Corresponding author: ]{tom@wenckebach.net}
\affiliation{PSI Center for Neutron and Muon Sciences, CH-5232 Villigen-PSI, Switzerland}

\author{Yifan Quan}\email[Corresponding author: ]{yquan@sas.upenn.edu}
\affiliation{Department of Chemistry, University of Pennsylvania, Philadelphia, PA 19104, USA}
\date{\today}

\maketitle

\tableofcontents

\section{Experimental}
The experiments were carried out on single crystals of naphthalene doped with pentacene using a home-built triplet DNP/EPR apparatus \cite{Eichhorn2013}. The setup consists of a laser system for efficient triplet excitation, an X-band pulsed DNP/EPR spectrometer with an electromagnet, two NMR spectrometers for measuring the nuclear polarization, and a helium flow cryostat with optical access for cooling the sample. More details on the setup and its components can be found in a recent review \cite{Hautle2024}.

\subsection{Pentacene-naphthalene single crystal sample}

\begin{figure}[htbp]
\centering
\includegraphics[scale=0.45]{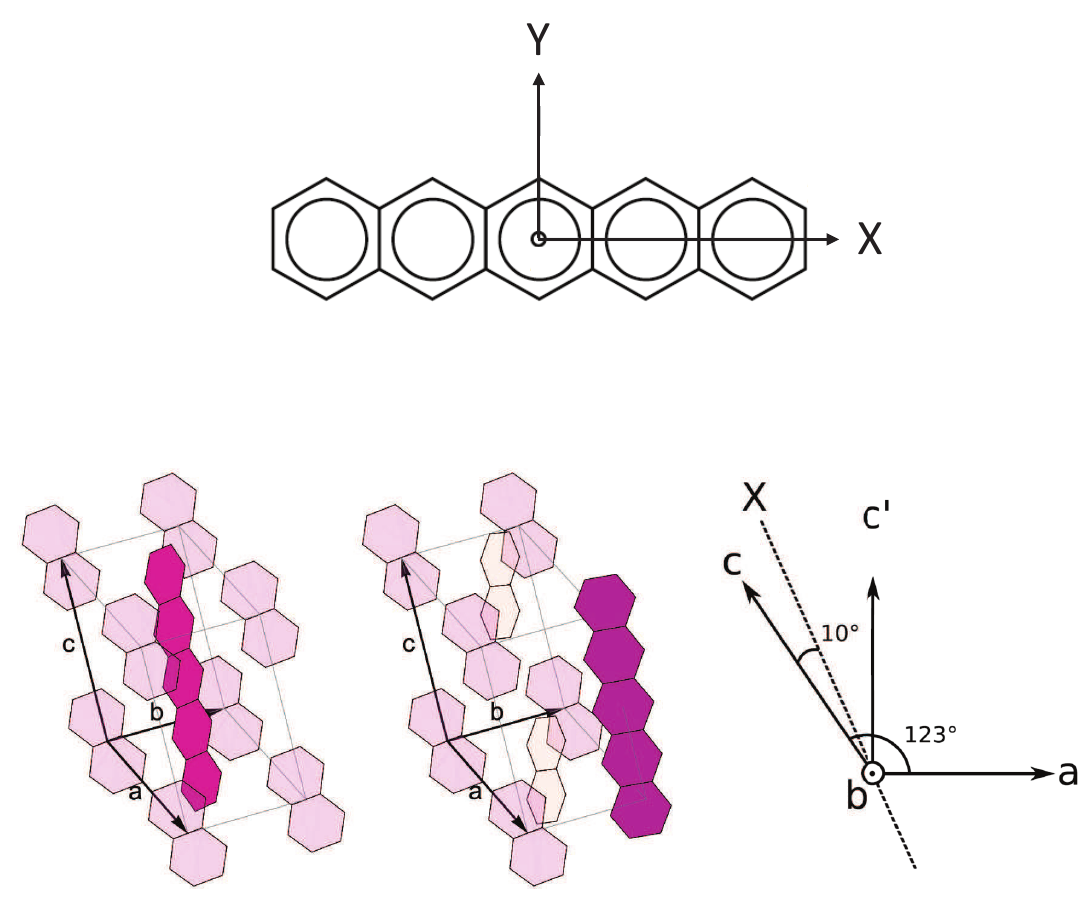}
\caption{\textbf{Pentacene molecule and the structure of pentacene:naphthalene crystal.} Top: Pentacene molecule with its principal axes $X$ and $Y$ (in plane) and $Z$ (normal to the molecular plane). Bottom: Crystal structure of pure naphthalene with lattice parameters {\em a} $\sim 8.2$~\AA, {\em b} $\sim 6.0$~\AA, {\em c} $\sim 8.7$~\AA, and $\beta \sim 123^\circ$ \cite{Brock:1982}. The crystal belongs to the space group C$_{2\lambda}^5$P2$_{1}/a$ \cite{Robertson:1933}. The $ab$-plane is the cleavage plane. Two naphthalene molecules can be replaced by a single pentacene molecule. Both substitution sites share the same pentacene $X$-axis, while the $Y$- and $Z$-axes are oriented at $47 \pm 3^\circ$. The $X$-axis lies in the $ac$-plane at an angle of $10^\circ$ relative to the $c$-axis \cite{van1980epr}.}
\label{fig:Crystal}
\end{figure}

Figure \ref{fig:Crystal} illustrates the molecular structures of naphthalene and pentacene, along with the convention for their principal $X$, $Y$ and $Z$-axes. Pure naphthalene crystallizes in a monoclinic structure belonging to the space group C$_{2\lambda}^{5}$P2$_{1}/a$. The $ab$-plane is the cleavage plane of the crystal, and the lattice parameter along the $c$-axis is comparable to the long molecular axis ($X$) of pentacene. It is possible to build a pentacene molecule into the crystal structure of naphthalene by replacing two naphthalene molecules by a single
pentacene molecule. When pentacene is incorporated into the lattice, two crystallographically inequivalent substitution sites are formed. These two orientations share a common molecular $X$-axis, while their respective $Y$- and $Z$-axes are tilted by $47 \pm 3^\circ$ relative to each other. The molecular $X$-axis lies in the $ac$-plane and is inclined by approximately $10^\circ$ with respect to the crystallographic $c$-axis~\cite{van1980epr}.

To minimize paramagnetic impurities and undesired spin defects, the naphthalene host material was extensively purified using zone-refinement techniques. Deuterated pentacene was customly synthesized and further purified by ISOTEC (Sigma-Aldrich). All sample preparation and handling were performed under a nitrogen atmosphere inside a glovebox to prevent oxidation and contamination. Large single crystals with volumes of several cm$^3$ were grown using a self-seeding vertical Bridgman technique. Detailed procedures for crystal growth, dopant incorporation, and post-growth characterization can be found in \cite {Hautle2024}.

The sample used in all the experiments discussed in the main text was cut out of a large single crystal of naphthalene doped with pentacene-$d_{14}$. Its size was $3.2 \times 4.2 \times 2.5$\,mm$^3$ ($a \times b \times c^\prime$), and its pentacene concentration was determined with optical transmission spectroscopy to be $(7.9\pm 0.5)\times 10^{-5}$\,mol/mol $(0.71 \pm 0.045$\,mM).
The sample was mounted on a polychlorotrifluorethylene (PCTFE) holder attached to a sample rod, introduced in a helium flow cryostat and subjected to a magnetic field of around 0.36\,T generated by an electromagnet. The magnetic field could be rotated in the horizontal plane, and the sample was oriented such that its $ac$-plane, containing the pentacene molecular $X$-axis, coincided with this plane. The sample rod itself could be rotated continuously over 360$^\circ$ and positioned with a precision of 0.2$^\circ$ using a laser pointer. By rotation of the magnet and/or the sample rod, the pentacene molecular $X$-axis could be precisely aligned parallel to the magnetic field.

\subsection{Triplet excitation}
The pentacene molecules were excited into the triplet state with an infrared diode-pumped disk laser (Jenlas disk IR50). Its output is frequency-doubled to 515\,nm using a lithium triborate (LBO) nonlinear crystal and coupled into a high numerical-aperture multimode fiber with a core diameter of 1.5 mm. The fiber is connected at the other end to a collimator located at the bottom of the cryostat. It focuses the unpolarized light propagating along the vertical $b$-axis of the naphthalene crystal to a waist of 7 mm at the sample position, ensuring homogeneous illumination.
The laser generates pulses of 400\,ns length at repetition rates of up to 8\,kHz. For the polarization build-up, a repetition rate of 1\,kHz was used, whereas for the EPR measurements 100\,Hz was chosen to minimize sample heating. The pulse energy at the entrance window of the cryostat was approximately 0.5\,mJ.

\subsection{EPR and DNP}
Pulsed EPR and DNP experiments were performed using a home-built X-band spectrometer with wide-bandwidth phase-sensitive detection. Pulse generation and synchronization are controlled by a Spincore PulseBlasterESR-Pro card, allowing the generation of microwave pulses as short as 12 ns. The microwave pulses are amplified by solid-state X-band amplifiers, and the demodulated signal is recorded using a high-speed digitizer. The EPR resonator, consisting of a sapphire dielectric ring inside a cylindrical cavity, provides a high filling factor and optical access for laser excitation. 

\begin{figure}[htbp]
\centering
\includegraphics[scale=0.5]{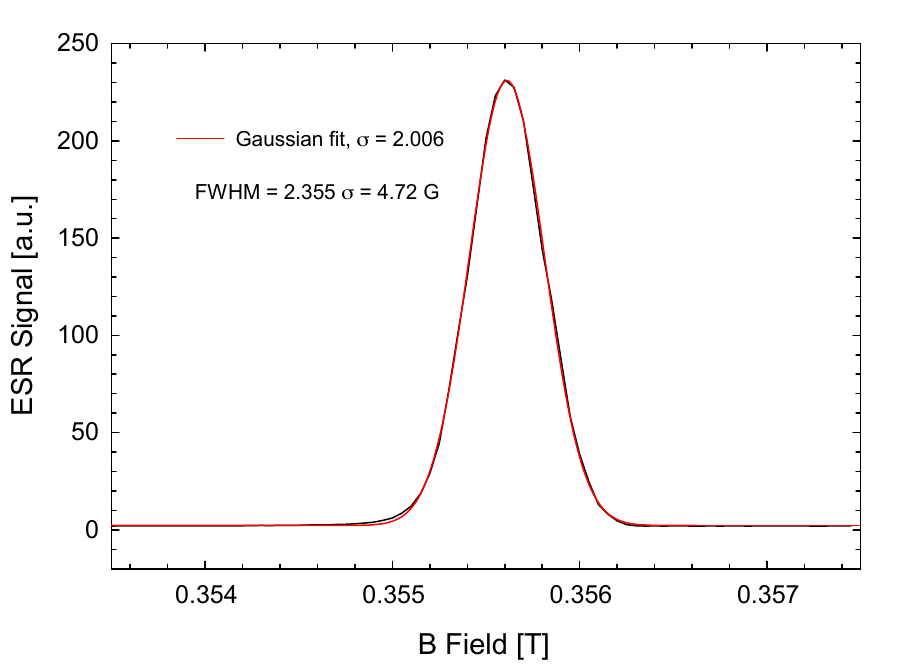}
\caption[EPR spectrum of the high field transition]
{\textbf{EPR spectrum of the high field transition.} EPR spectrum of the high-field transition of the pentacene triplet state with the $X$-axis aligned parallel to the magnetic field, obtained from pulsed EPR measurements at a fixed frequency of 9.437\,GHz as a function of magnetic field at $T = 80$\,K. The spectrum is fitted with a Gaussian of width $\sigma = 0.200$\,mT, corresponding to a FWHM of $2\sqrt{2\ln 2}\sigma = 0.471$\,mT.}
\label{fig:EPRspectrum}
\end{figure}

The spectrometer is synchronized to the laser pulses via a fast photodiode detecting the light reflected from the first focusing lens after the laser output. 
Following laser excitation, a standard Hahn-echo microwave sequence was employed to measure the triplet EPR signal. The microwave frequency was fixed at approximately 9.44\,GHz, while the magnetic field was tuned to the $T_0 \leftrightarrow T_-$ high-field transition at approximately 0.355\,T. The optimum, i.e., parallel orientation of the crystal $X$-axis with respect to the magnetic field was determined by minimizing the width of the EPR spectrum while maximizing the EPR signal. Figure \ref{fig:EPRspectrum} shows the optimized EPR spectrum.

\begin{figure}[htbp]
\centering
\includegraphics[scale=0.5]{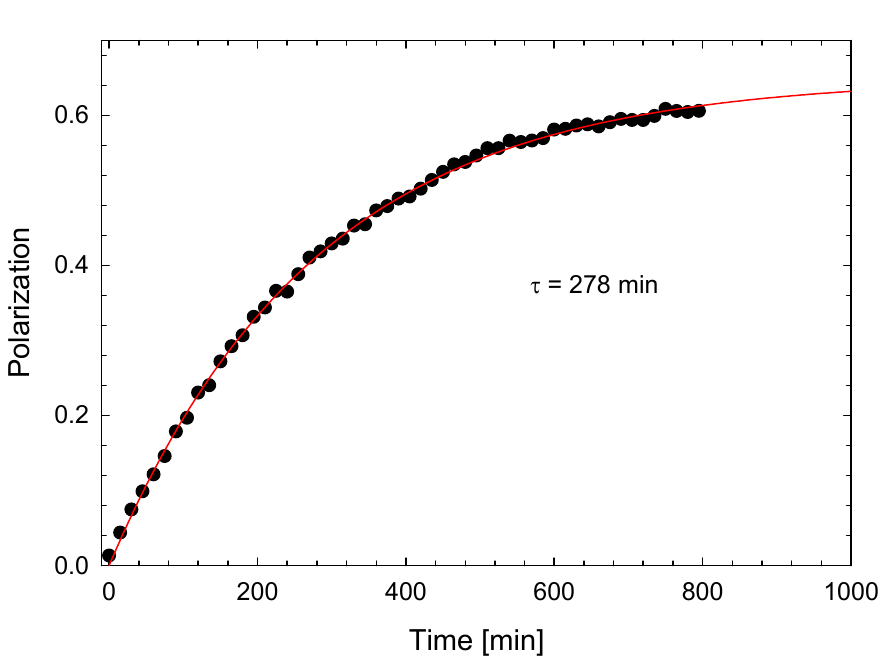}
\caption[Proton polarization build-up via ISE]
{\textbf{Proton polarization build-up via triplet-DNP.}Proton polarization build-up with ISE at a repetition rate of 1\,kHz in the naphthalene:pentacene crystal at 25\,K and 0.355\,T. The polarization was monitored by pulse NMR, which was cross calibrated against the CW-NMR.}
\label{fig:ISE_buildup}
\end{figure}

Fast magnetic field sweeps required for polarization build-up via the integrated solid effect (ISE) \cite{Henstra1990,Henstra2014,Eichhorn2014,Quan2022,Hautle2024} are generated by a 10-turn saddle coil wound on a Teflon holder around the resonator and driven by a linear high-current amplifier, enabling current sweeps of up to $\pm15$\,A within 15\,$\mu$s, corresponding to field sweep rates of up to 0.6\,mT/$\mu$s. The current sweep waveform is provided by a preprogrammed arbitrary waveform generator, triggered by the EPR system, and can be operated at repetition rates of up to 4\,kHz. In the experiments presented, the sample was polarized at a repetition rate of 1\,kHz. Synchronized with the laser pulse, a 7\,$\mu$s microwave pulse at approximately 9.44\,GHz was applied while the magnetic field was adiabatically swept at a rate of 0.32\,mT/$\mu$s across the 0.47\,mT (FWHM) inhomogeneously broadened high-field transition at 0.355\,T, with the flow cryostat operated at 25\,K. Figure \ref{fig:ISE_buildup} shows the build-up of the proton polarization in naphthalene, monitored by pulse NMR, which rises with a time constant of $\tau = 273$\,min to a maximum of approximately $P = 0.6$.

\subsection{NMR and polarization measurement}
The proton polarization was measured using a pulse NMR spectrometer based on a Spincore RadioProcessor card and a CW-NMR system based on a Q-meter \cite{Court1993,Court2004}. The former is used with a 5-turn solenoid coil located just below the sapphire dielectric ring and a matching circuit positioned just above the EPR resonator. It serves to monitor the polarization during DNP by means of small angle tipping pulses. The CW-NMR system is operated with a 15-turn solenoid of 14\,mm length and 8\,mm inner diameter, located about 31\,mm above the center of the resonator. For measurements, the sample is lifted into the center of the coil, thus obtaining a better filling factor and higher sensitivity. It serves for calibration of the polarization, and to eliminate temperature dependencies, its matching and tuning circuits are positioned outside the cryostat. The vertical position of the sample can be adjusted with a precision of $\pm 0.05$\,mm by means of a caliper. The CW-NMR system has been calibrated by determining the absolute value of the proton spin polarization $P$ of a sample from a neutron transmission measurement.
Neutron scattering is strongly spin-dependent, with neutrons anti-parallel to the proton spins scattered much more than those with parallel spin \cite{Lushchikov1969,Haag2012}, and the polarization-dependent cross-section is well documented in the literature \cite{Quan2019}. By measuring the transmission of a fully polarized neutron beam through the sample with spins parallel and anti-parallel to the magnetic field for some 10\,s each, the nuclear polarization can be determined with an accuracy of approximately 3\%, providing a reliable calibration for the NMR system.

\begin{figure}[htbp]
\centering
\includegraphics[scale=0.5]{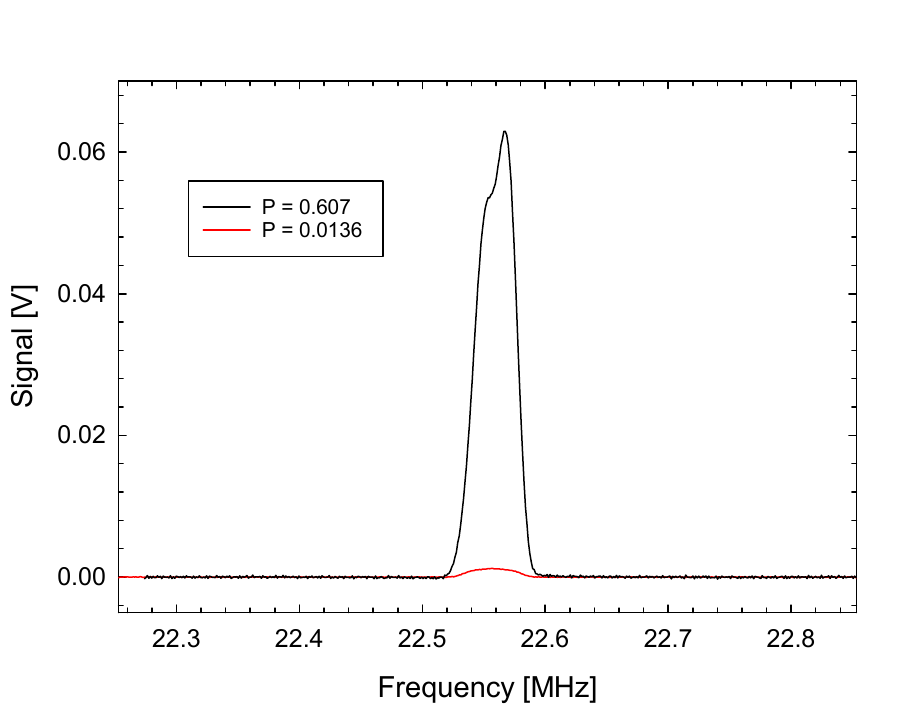}
\caption{\textbf{CW-NMR Signals.} CW-NMR signals of the polarized naphthalene crystal taken at a magnetic field of $B_0 = 0.52$\,T.}
\label{fig:CW-NMR}
\end{figure}

\subsection{Protocol for echo decay measurements}
To keep systematic errors under control, the spin-echo coherence time $T_2$ as a function of proton polarization was measured at 80 K in a single series, starting at the highest polarization after build-up and then at progressively reduced polarization down to thermal equilibrium.

For each polarization value, the triplet electron spin $T_2$ was measured twice using a Hahn-echo sequence with a $\pi/2$ pulse of 24\,ns, incrementing the inter-pulse delay $\tau$ from 1.2 to 10\,$\mu$s in 880 steps of 10\,ns, with 100 averages recorded at each point. The minimum value of $\tau$ was limited by the ringing of the resonant circuit. The polarization was measured before and after each sequence using CW-NMR. The nuclear polarization loss caused by the laser irradiation and microwave pulsing at 100\,Hz, was approximately 2\,\% during the 2 times 15\,min (total 30 min) long measurements, corresponding to $T_1 \sim 25$\,h ($\sim 9\times10^6$ Hahn echo cycles).

\begin{figure}[htbp]
\centering
\includegraphics[scale=0.5]{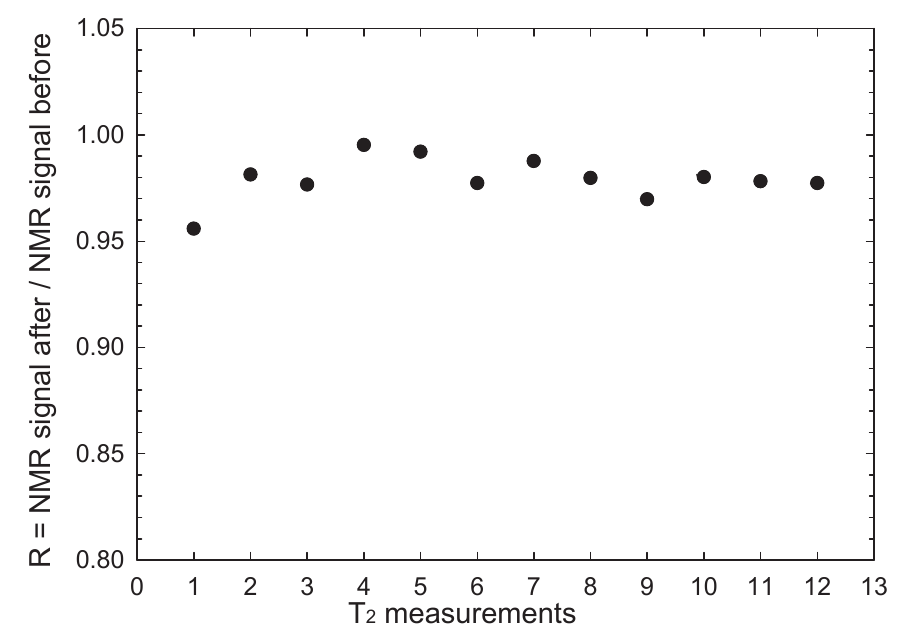}
\caption{\textbf{Nuclear Polarization Loss During Laser Irradiation and Microwave Pulsing.}
Measured nuclear polarization loss caused by the laser irradiation and microwave pulsing at 100~Hz, was approximately 2~\% during the 2 times 15~min (total 30 min) long measurements.
}
\label{fig:2sample}
\end{figure}

\subsection{Comparison between different samples}

A second sample, grown in a separate batch, was measured and reproduces the observed behavior, confirming the robustness of our results.

\begin{figure}[htbp]
\centering
\includegraphics[scale=0.25]{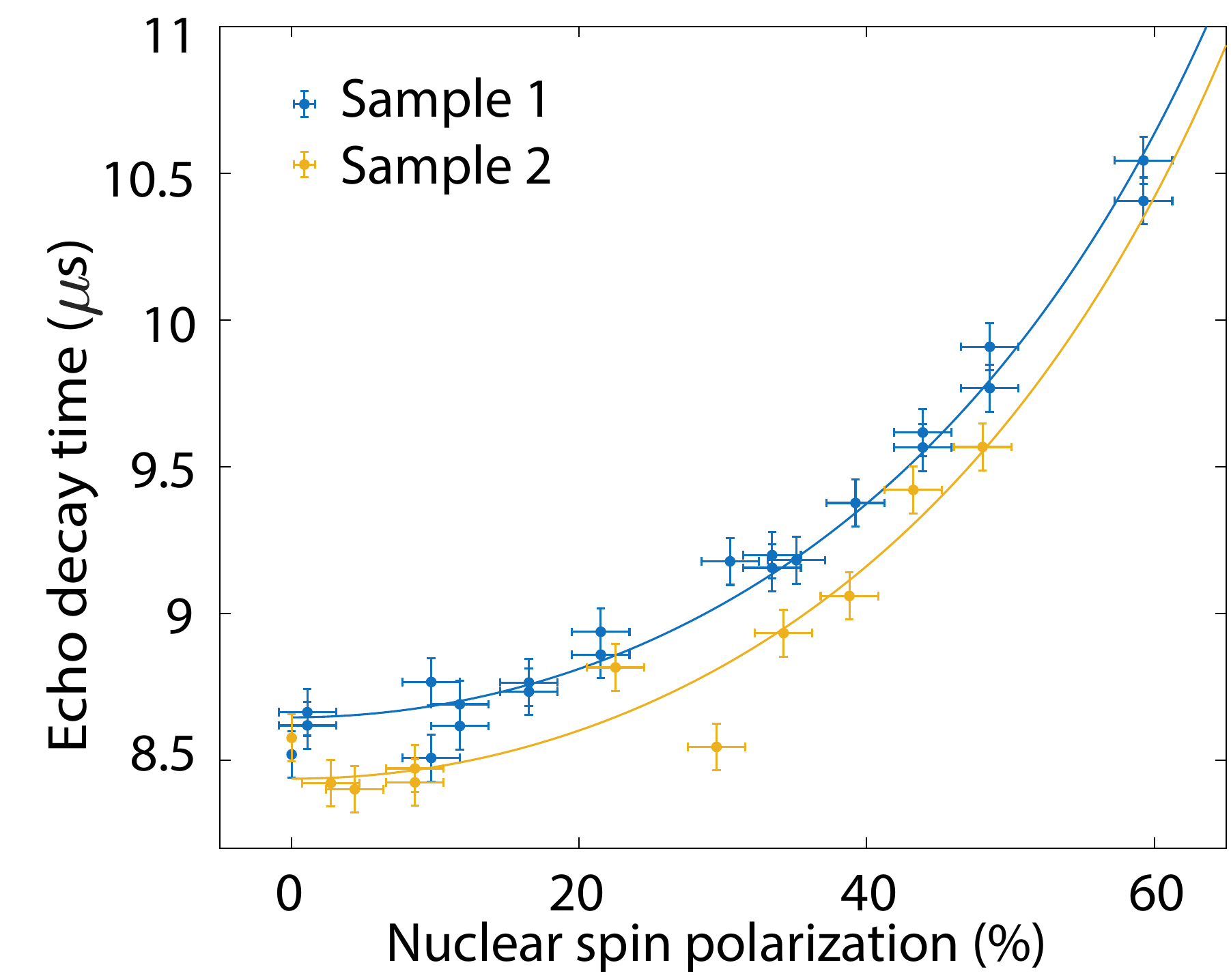}
\caption{\textbf{Coherence time for two independently grown samples.}
Extracted coherence time $T_2$ as a function of nuclear polarization $P_I$ for two independently grown samples, showing consistent behavior across batches.
}
\label{fig:2sample}
\end{figure}

\section{Spin-Echo Decoherence under a Fluctuating Magnetic Field}

In this appendix, we briefly derive the scaling relation between the spin-echo coherence time $T_2$ and the fluctuating magnetic field generated by a spin bath. We consider a central spin interacting with a stochastic magnetic field $B(t)$ along the longitudinal direction:
\begin{equation}
H(t)=\gamma S_z B(t),
\end{equation}
where $\gamma$ is the gyromagnetic ratio of the central spin. The fluctuating magnetic field is characterized by its autocorrelation function and power spectral density (PSD):
\begin{equation}
    G(|t_1-t_2|) \equiv \langle B(t_1) B(t_2)\rangle,
\end{equation}
\begin{equation}
S(\omega) = \int_{-\infty}^{\infty}\mathrm{d}t\,\mathrm{e}^{-i\omega t} G(t).
\end{equation}
For a Hahn spin-echo sequence with a $\pi$ pulse applied at time $\tau$, the coherence of the central spin is determined by the accumulated stochastic phase:
\begin{equation}
\langle S_x(t)\rangle
=
\left\langle
\exp\left[-i\gamma\int_0^{t} \mathrm{d}t'\, B(t') f(t')\right]
\right\rangle,
\end{equation}
where $S_x(t)$ is the experimentally measured signal and $f(t)$ is the modulation function defined by the pulse sequence:
\begin{equation}
f(t)=
\begin{cases}
1, & t\in[0,\tau),\\
-1, & t\in[\tau,2\tau].
\end{cases}
\end{equation}

Assuming that $B(t)$ is a stationary, zero-mean Gaussian process, the coherence decays as
\begin{equation}
\begin{aligned}
\langle S_x(t)\rangle
&=
\exp\left[
-\gamma^2 \int_0^{2\tau}\mathrm{d}t_1 \int_0^{2\tau}\mathrm{d}t_2\, 
f(t_1)f(t_2)\langle B(t_1)B(t_2)\rangle
\right] \\
&=
\exp\left[
-\frac{\gamma^2}{\pi}
\int_0^\infty \mathrm{d}\omega\, S(\omega)\, |F(\omega)|^2
\right],
\end{aligned}
\label{eqS:decay}
\end{equation}
where $|F(\omega)|^2$ is the filter function for the Hahn echo sequence:
\begin{equation}
|F(\omega)|^2 \equiv 
\frac{\left|\int_0^{2\tau}\mathrm{d}t\, f(t)\, \mathrm{e}^{-i\omega t}\right|^2}{\omega^2}
=
\frac{16}{\omega^2}\sin^4\!\left(\frac{\omega \tau}{2}\right).
\end{equation}
When the fluctuating magnetic field is generated by a nuclear spin bath, it is commonly modeled as an Ornstein-Uhlenbeck (OU) stochastic process, leading to a Lorentzian power spectral density (PSD). In the presence of Larmor precession, the PSD can be approximated as
\begin{equation}
S(\omega)\approx
\frac{2B_{\mathrm{rms},z}^2\tau_c}{1+\omega^2\tau_c^2}
+
\frac{B_{\mathrm{rms},\perp}^2\tau_c}{1+(\omega-\omega_{0I})^2\tau_c^2}
+
\frac{B_{\mathrm{rms},\perp}^2\tau_c}{1+(\omega+\omega_{0I})^2\tau_c^2},
\label{eq:spec_full}
\end{equation}
where $\omega_{0I}$ is the Larmor frequency of the nuclear spins under an external magnetic field, and $\tau_c$ is the correlation time of the bath. The quantities $B_{\mathrm{rms},z}$ and $B_{\mathrm{rms},\perp}$ characterize the root-mean-square amplitudes of the longitudinal and transverse components of the fluctuating field, respectively. These arise from nuclear spin flip-flop dynamics and coherent Larmor precession.

The effective magnetic field can be written as
\begin{equation}
B(t)=\sum_j \left[ A_{zz}^{(j)} I_z^{(j)}(t) + A_{z\perp}^{(j)} I_\perp^{(j)}(t)\right],
\end{equation}
where $A_{zz}^{(j)}$ and $A_{z\perp}^{(j)}$ are the longitudinal and transverse hyperfine coupling components to the $j$th nuclear spin.

From the definition of the PSD and its relation to the autocorrelation function, the variance of each component is
\begin{equation}
\begin{aligned}
B_{\mathrm{rms},\sigma}^2
&= \langle B_\sigma(t) B_\sigma(0)\rangle \\
&= \sum_{j,k} A_{z\sigma}^{(j)} A_{z\sigma}^{(k)}
\Big[
\langle I_\sigma^{(j)}(t) I_\sigma^{(k)}(0) \rangle
-
\langle I_\sigma^{(j)}\rangle \langle I_\sigma^{(k)}\rangle
\Big],
\end{aligned}
\end{equation}
where $\sigma = z,\perp$. Assuming near-uncorrelated nuclear spins, this reduces to
\begin{equation}
B_{\mathrm{rms},\sigma}^2
\approx 
\sum_j (A_{z\sigma}^{(j)})^2
\Big[
\langle I_\sigma^{(j)}(t) I_\sigma^{(j)}(0) \rangle
-
\langle I_\sigma^{(j)}\rangle^2
\Big].
\end{equation}

For a partially polarized nuclear spin-$1/2$ bath, we take
$\langle I_z^{(j)}\rangle = P_I$, $\langle I_\perp^{(j)}\rangle = 0$, which yields
\begin{equation}
B_{\mathrm{rms},z}^2
\approx \sum_j \frac{1}{4}(A_{zz}^{(j)})^2\left(1-P_I^2\right), 
\qquad
B_{\mathrm{rms},\perp}^2
\approx \sum_j \frac{1}{4} (A_{z\perp}^{(j)})^2 .
\label{eqS:brms_pol}
\end{equation}

Under typical experimental conditions, the spin-echo evolution is slow compared to the nuclear Larmor frequency, such that the filter function (Eq.~\eqref{eq:spec_full}) is strongly peaked near zero frequency. As a result, according to Eq.~\eqref{eqS:decay}, the echo signal is primarily sensitive to the low-frequency component of the noise spectrum, while the contributions at $\omega \approx \pm \omega_{0I}$ (e.g., $\sim 15~\mathrm{MHz}$) are effectively suppressed.

There are two relevant limits set by the relation between the bath correlation time $\tau_c$ and the echo evolution time $\tau$.
In the fast-bath limit, $\tau_c \ll \tau$, the noise spectrum is broad compared to the filter function. The filter function effectively samples the PSD near zero frequency, yielding
\begin{equation}
\langle S_x(t)\rangle
\approx 
\exp\left[-\gamma^2 B_{\mathrm{rms},z}^2 \tau_c \cdot 2\tau \right],
\end{equation}
which corresponds to exponential decoherence with a rate proportional to $B_{\mathrm{rms},z}^2 \tau_c$.

In the opposite slow-bath limit, $\tau_c \gg \tau$, the noise is quasi-static during the echo sequence, and the PSD becomes sharply peaked near zero frequency. In this case, the coherence decay is determined by the low-frequency expansion of the filter function, giving
\begin{equation}
\langle S_x(t)\rangle
\approx 
\exp\left[-\frac{\gamma^2 B_{\mathrm{rms},z}^2}{12\tau_c } \cdot(2\tau)^3 \right],
\end{equation}
which leads to a stretched (cubic) exponential decay.

In our experimental regime, the system operates close to the slow-bath limit. In both limits, the dependence of the dephasing time on nuclear spin polarization arises from the $B_{\mathrm{rms},z}$ term in Eq.~\eqref{eqS:brms_pol}, as well as from the correlation time $\tau_c$, which will be discussed below.

As mentioned in the main text, the correlation time of the nuclear spin bath is primarily governed by spin-spin interactions:
\begin{equation}
    \frac{1}{\tau_c} = \Gamma_{s\text{-}l} + \Gamma_{f\text{-}f} \approx \Gamma_{f\text{-}f},
\end{equation}
where $\Gamma$ denotes the decay rate, and $s$-$l$ and $s$-$s$ represent spin-lattice and spin-spin flip-flop interactions, respectively. The correlation time is set by nuclear spin flip-flop processes, which are directly related to the homogeneous linewidth of the nuclear spins. This linewidth has been derived analytically in Refs.~\cite{Abragam1973,Abragam1961}, yielding a square-root dependence on the nuclear spin polarization:
\begin{equation}
\tau_c \propto J_{\mathrm{rms}}^{-1}(1-P_I^2)^{-1/2},
\end{equation}
where $J_{\mathrm{rms}}$ characterizes the typical off-diagonal nuclear-nuclear dipolar coupling.

Here we provide an alternative perspective based on Fermi's golden rule. The decay rate induced by nuclear spin flip-flop processes can be expressed as
\begin{equation}
    \Gamma_{f\text{-}f} = J_{\mathrm{rms,\perp}}^2\,\tilde{S}(\omega_{0I}),
    \label{eqS:fgr}
\end{equation}
where $\tilde{S}(\omega)$ is the effective spectral density for flip-flop processes. Since flip-flops occur only between oppositely polarized spins, the available pair population introduces a factor $\frac{1-P_I}{2}\cdot\frac{1+P_I}{2}$, giving
\begin{equation}
    \tilde{S}(\omega) \propto (1-P_I^2)\, S(\omega).
\end{equation}

The nuclear spin spectral density is approximated by a Lorentzian centered at the Larmor frequency:
\begin{equation}
    S(\omega) = \frac{\Gamma/2}{(\omega-\omega_{0I})^2 + (\Gamma/2)^2},
\end{equation}

where $\Gamma$ is the intrinsic linewidth. Evaluating Eq.~\eqref{eqS:fgr} at $\omega=\omega_{0I}$ yields
\begin{equation}
    \Gamma_{f\text{-}f} \propto J_{\mathrm{rms,\perp}}^2 (1-P_I^2)\, \frac{1}{\Gamma}.
\end{equation}


Assuming self-consistently that the linewidth is determined by the flip-flop processes and inhomogeneous broadening, i.e., $\Gamma \sim \Gamma_{f\text{-}f}+\Gamma_{z}$, where $\Gamma_z\sim\sqrt{\sum J_{zz}^2\mathrm{Var}[I_z]}$, we obtain:
\begin{equation}
    \Gamma_{f\text{-}f} \propto J_{\mathrm{rms,\perp}}^2 (1-P_I^2)\, \frac{1}{\Gamma_{f\text{-}f}+J_{rms,z}\sqrt{1-P_I^2}}.
\end{equation}
Solving the self-consistent equation yeilds:
\begin{equation}
\begin{aligned}
  &\Rightarrow  \Gamma_{f\text{-}f} \propto \sqrt{1-P_I^2}.
\end{aligned}
\end{equation}


\section{CCE simulation}

\begin{figure}[htbp]
\centering
\includegraphics[width=1\linewidth]{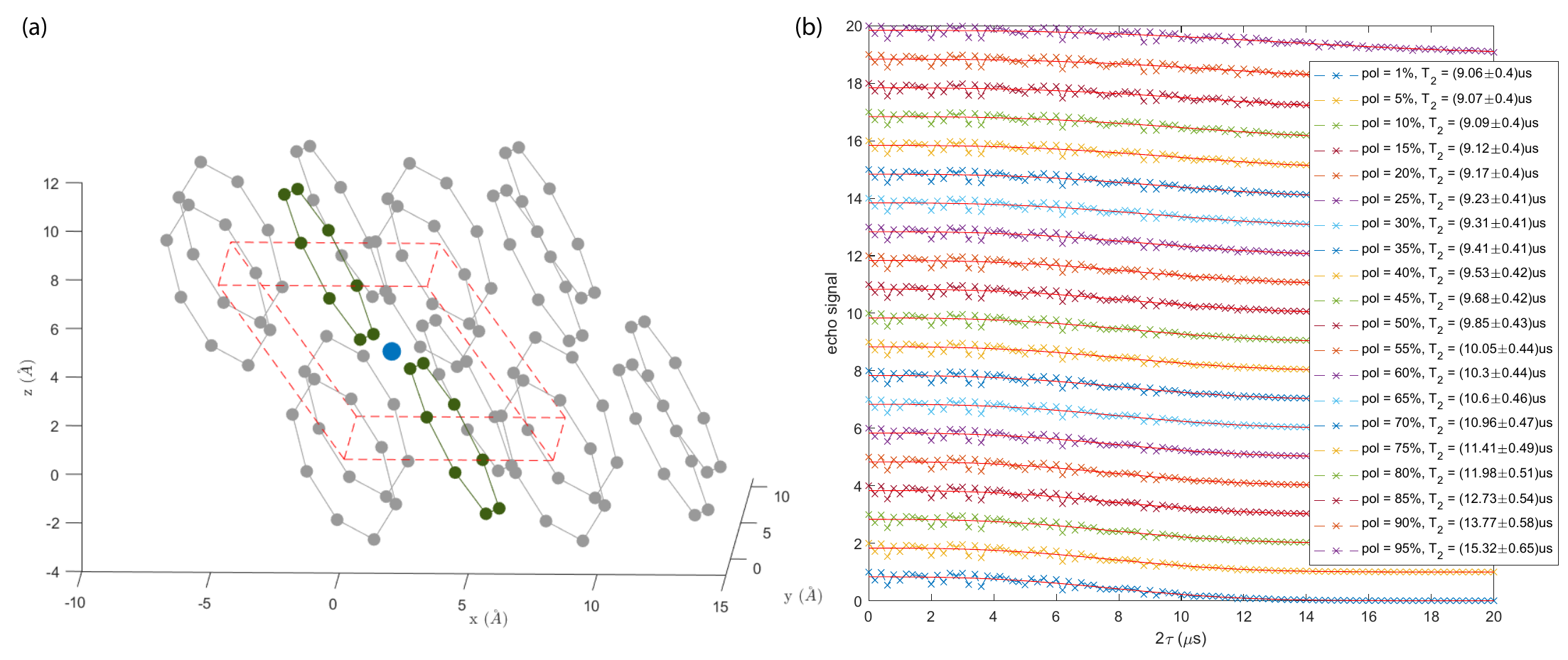}
\caption{\textbf{Nuclear spin distribution and raw CCE data.} 
(a) Proton spin positions (gray and green dots) in the naphthalene crystal lattice. A unit cell is highlighted by the red dashed outline. Two naphthalene molecules at one of the quivalent sites (green) can be substituted by a pentacene molecule, with the electron spin (blue dot) assumed to be localized at its center. 
(b) Raw CCE simulation data used in the main text, together with the corresponding fits.}
\label{figS:cce}
\end{figure}

Cluster correlation expansion (CCE) simulations are performed using the PyCCE package~\cite{onizhuk2021pycce}. We first construct the lattice structure of the naphthalene crystal. Figure~\ref{figS:cce} (a) shows the positions of the proton spins. There are two equivalent molecular sites in each unit cell, and pentacene substitutes for two naphthalene molecules at either site. In the simulations presented in the main text, the electron spin is assumed to be localized at the center of the pentacene molecule. The initialization nuclear spin state is set to be:
\begin{equation}
    \rho_n = \bigotimes_{(j)}\frac{1}{2}(\mathbb{I}+P_I\sigma_z^{(j)}),
\end{equation}
where $\mathbb{I}$ is the identity operator, $\sigma_z^{(j)}$ is the Pauli operator for $(j)^{th}$ nuclear spin. The raw simulated data, together with stretched-exponential fits supporting the main text, are shown in Fig.~\ref{figS:cce} (b).

The coupling between the electron spin and the nuclear spins is calculated based on the magnetic dipolar interaction. Within the central spin model, the nuclear spin dynamics are conditioned on the state of the electron spin. We tested convergence with respect to the cluster order, the bath cutoff radius $r_{\mathrm{bath}}$, and the dipolar cutoff radius $r_{\mathrm{dipole}}$ (which determines whether two nuclear spins are considered coupled), for different nuclear spin polarization conditions. These results demonstrate that CCE-2 with $r_{\mathrm{bath}} = 40~\text{\AA}$ and $r_{\mathrm{dipole}} = 6~\text{\AA}$ provides sufficient convergence across all polarization values, consistent with previous CCE studies on nuclear spin bath~\cite{onizhuk2021probing}.

\begin{figure}[htbp]
\centering
\includegraphics[width=1\linewidth]{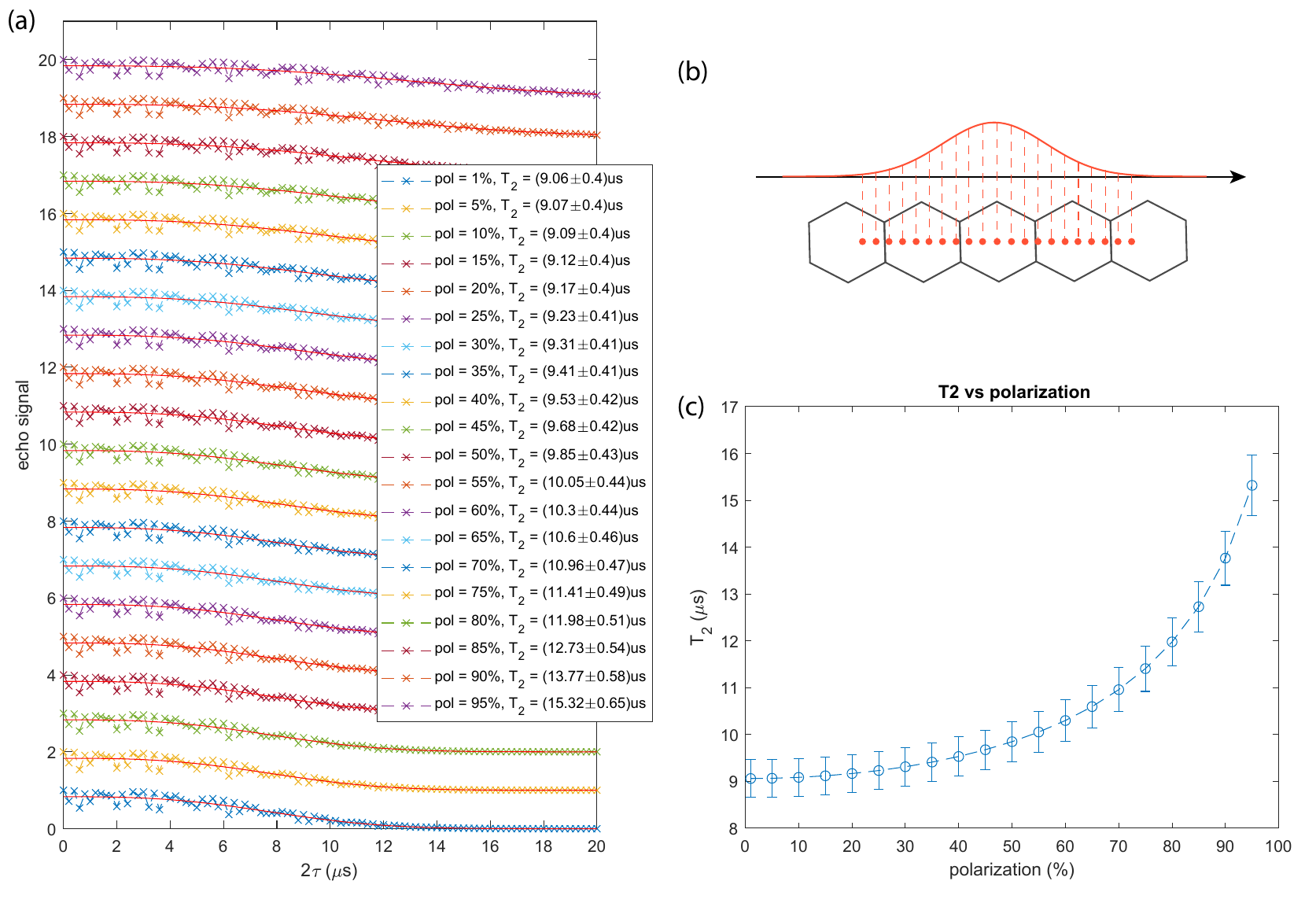}
\caption{\textbf{CCE simulation with spatially distributed electron spin.} 
(a) Hahn-echo signals obtained from CCE simulations, averaged over different electron spin positions. 
(b) Electron spin positions used in the ensemble average. The spatial distribution is approximated by a Gaussian profile over the pentacene molecule~\cite{richert2017delocalisation} and used as weighting for normalization. 
(c) Extracted coherence time $T_2$ from the averaged data.
}
\label{figS:cce_loc}
\end{figure}

 Since the electronic spin is in reality delocalized over the molecule, we additionally perform ensemble-averaged simulations by sampling its position at 11 locations along the pentacene molecular axis with displacements relative to the center given by 
 \[\{0,\,\pm0.5~\text{\AA},\,\pm1~\text{\AA},\,\pm1.5~\text{\AA},\,\pm2~\text{\AA},\,\pm2.5~\text{\AA},\,\pm3~\text{\AA},\,\pm3.5~\text{\AA},\,\pm4.5~\text{\AA},\,\pm4.5~\text{\AA},\,\pm5.5~\text{\AA}\},\]
 the result. In Fig.~\ref{figS:cce_loc}, we present the ensemble-averaged results over different electron spin locations, together with stretched-exponential fits. The extracted coherence time $T_2$ and stretching exponent $\mu$ as functions of nuclear spin polarization are also shown. We find that $T_2$ is only weakly dependent on the electron spin position, consistent with the assumption that dephasing is dominated by the large number of distant proton spins rather than strongly coupled nearby spins. 

\section{Electron Spin-Spin Interaction and Nuclear Spin Diffusion}

In our crystals the pentacene number density is typically in the order of $N=4 \times 10^{17} \mathrm{~cm}^{-3}$ so the distance between the pentacene molecules is roughly $l=\sqrt[3]{1 / N} \sim 13 \mathrm{~nm}$. This corresponds to an electron-electron dipolar interaction strength of approximately $20~\mathrm{kHz}$, representing an upper bound assuming all pentacene molecules are in the triplet state. Thus, decoherence arising from electron-electron interactions is expected to occur on a timescale much longer than the measured $T_2$ and can therefore be neglected, consistent with the experimental observations.

The spin diffusion time is given by $t=\frac{r^2}{4 D}$, where $r$ is the distance and $D$ is the spin diffusion constant, which is in the order of $8 \times 10^{-12} \mathrm{~cm}^2 \mathrm{~s}^{-1}$. Spin diffusion covers a distance of $l / 2\sim6.5 \mathrm{~nm}$ in about 1.3 ms, which is comparable to or faster than our typical pulse repetition time. Thus, spin diffusion is sufficiently fast to spread the polarization
between pentacene molecules, generating a homogeneous nuclear spin-bath polarization in the vicinity of the pentacene molecules after DNP as well as after each sensing pulse sequence.

\newpage\clearpage
\bibliographystyle{apsrev4-2}
\bibliography{PentaceneT2_NuclearPolarization_ref}